\documentstyle[prl,aps,twoside,floats,epsf]{revtex}

\begin{document}
\draft

\twocolumn[\hsize\textwidth\columnwidth\hsize\csname
  @twocolumnfalse\endcsname
  \draft   \title{Colossal  magnetooptical conductivity in doped
  manganites} 
  \author{ A.S. Alexandrov$^{1,*}$ and A.M. Bratkovsky$^{2,\dagger}$}
  \address{$^{1}$Department of Physics, Loughborough University,
  Loughborough LE11 3TU, UK\\
  $^{2}$Hewlett-Packard Laboratories, 3500~Deer~Creek
  Road, Palo Alto, California 94304-1392 }
\date{March 10, 1999}
\maketitle
\begin{abstract}
We show that the current carrier density collapse 
in doped manganites, which results from  bipolaron formation
in the  paramagnetic phase,   leads to a colossal change of the
optical conductivity in an external magnetic field at temperatures
close to the ferromagnetic transition.   As with the
colossal magnetoresistance (CMR)  itself, the corresponding 
magnetooptical effect
is explained by the dissociation of localized bipolarons  into mobile
polarons  owing to the exchange interaction with the localized Mn spins in the
ferromagnetic phase. The effect is positive at low frequencies and
negative in the high-frequency region. The present results agree
with available experimental observations.

\end{abstract}
\vskip 2pc ] 
\narrowtext
 As we have recently shown \cite{alebra,alebra2}, the interplay of the
electron-phonon and exchange interactions results in a current carrier
density collapse (CCDC) at the paramagnetic-ferromagnetic transition in
doped manganites. Owing
to the strong electron-phonon interaction, polaronic carriers are bound
into
almost immobile bipolarons in the paramagnetic phase. A few thermally
excited non-degenerate polarons polarize localized Mn $d$ electrons. As a
result, the exchange interaction breaks bipolarons below $T_{c}$ if the
$p-d$
exchange energy $J_{pd}S$ of the polaronic carriers with the localized Mn
$d$
electrons is larger than the bipolaron binding energy $\Delta $. Hence, the
density of current carriers (polarons) suddenly increases below $T_{c}$,
which explains the resistivity peak and CMR experimentally observed in many
ferromagnetic oxides \cite{van,helmolt,jin} as well as the giant
isotope effect \cite{mul,franck}, the tunneling gap \cite{tun}, the
specific
heat anomaly \cite{ram}, along with the temperature dependence of the dc
resistivity \cite{sch}.

Another conspicuous general feature of the doped manganites is the massive
spectral weight transfer in the optical conductivity  to lower frequencies
as the CMR materials are
cooled below the Curie temperature \cite{oki1,oki2,kim,ish}. We have
shown recently that this effect can be also understood as a result of CCDC
\cite{abopt}.
The high-temperature  optical conductivity is well described by
the bipolaron  absorption, while  the low temperature mid-infrared band
is
due to absorption by single  polarons.  
A theoretical model in which the breaking of bipolarons caused by
magnetic interactions below $T_c$ shifts the spectral weight  from the
bipolaronic peak to the polaronic one
is in quantitative agreement with the optical spectra of the layered
ferromagnetic ($T_c=125$K)  crystals  La$_{2-2x}$Sr$_{1+2x}$Mn$_2$O$_7$
\cite{ish} over the entire observed frequency and  temperature range.

In the present paper we show that the optical properties of doped manganites
should be extremely sensitive
to the magnetic field (as suggested by the CMR itself). We predict an increase
  of the optical conductivity at low frequencies and a decrease at high
  frequencies with increasing magnetic field. The relative change of
the optical   conductivity is found to be  very large near the phase
transition. 

The optical intraband conductivity of a charge-transfer doped insulator
with
(bi)polaronic carriers is the sum of the polaron $\sigma _{p}(\nu )\equiv
en\mu_{p}(\nu)$ and
bipolaron $\sigma _{b}(\nu )\equiv e(x-n)\mu_{b}(\nu)$ contributions at a
given frequency $\nu $
\cite{abopt}.
Their frequency dependence is described in the literature (for
references see \cite
{bri,mah,alemot,abopt}). The polaron and bipolaron  optical mobilities
have almost a Gaussian shape given by
\begin{equation}
\mu _{p,b}(\nu ) ={\mu _{0}{\cal T}^{2}\over{\nu \gamma _{p,b}}}
\exp \left[ -(\nu -\nu _{p,b})^{2}/\gamma
_{p,b}^{2}\right],     \label{eq:sigma}
\end{equation}
where $\mu _{0}=2\pi ^{1/2}e/a$ is a constant with $a$ the lattice
spacing, ${\cal T}$ the hopping integral, $n=n(T,H)$ the (atomic) polaron
density for a given temperature $T$ and magnetic field $H$, and $x$ the
doping level. Here and below we assume $\hbar =c=1$.
In general  the polaron and bipolaron absorption have different maxima
located around $\nu _{p}=4E_{a}$ for polarons, and around
$\nu _{b}=2E_{p}-V_{c}$  for bipolarons \cite{abopt}, where $E_{a}$ is the
dc activation energy, $E_{p}$ is the polaron (Franck-Condon) level shift
and $V_{c}$ is the (inter-site) Coulomb repulsion. The
broadenings of the polaronic, $\gamma_p$, and bipolaronic, $\gamma_b$,
absorption are also different.

At temperatures above the ferromagnetic transition  the polaron
density is very low owing to CCDC \cite{alebra}, so the intraband
conductivity is due to bipolarons alone,
\begin{equation}
\sigma (\nu )={\frac{e\mu _{0}x{\cal T}^{2}}{{\nu \gamma _{b}}}}\exp
\left[
-(\nu -\nu _{b})^{2}/\gamma _{b}^{2}\right].
\end{equation}
This expression fits the experiment\cite{ish} fairly well with $\nu
_{b}~=~1.24$ eV and $\gamma _{b}~=~0.6$ eV \cite{abopt}.
When the temperature drops below $T_{c}$, at least some of the bipolarons
are broken apart by the exchange interaction with Mn sites, because one of the
spin-polarized polaron bands falls suddenly below the bipolaron level by an
amount $(J_{pd}S-\Delta )/2$ \cite{alebra}. The intraband optical
conductivity is determined now by both the polaronic and bipolaronic
contributions,  which explains the sudden spectral
weight transfer from $\nu =\nu _{b}$ to $\nu =\nu _{p}$ observed below
$T_{c}$ in the ferromagnetic manganites \cite{oki2,kim,ish}.
 The experimental spectral shape in La$_{2-2x}$Sr$_{1+2x}$Mn$_2$O$_7$
 at $T=10$K is well described
  with
 $n=x/5$, $\nu_{p}~=~0.5$ eV and $\gamma_{p}~=~0.3$ eV \cite{abopt}.


The presence of a magnetic field changes the balance between polarons
and bipolarons, 
resulting in a change of the optical conductivity. To quantify this
$magnetoptical$ effect we introduce the relative magnetoconductivity as
\begin{equation}
\delta\sigma /\sigma \equiv [\sigma(\nu,H) - \sigma(\nu,0)]/\sigma(\nu,0).
\label{eq:ds}
\end{equation}
Carriers in manganites are heavy. Hence, we do not expect any orbital
effect of the magnetic field on the mobility. Only the carrier density is
sensitive to the field in our theory. As a result we obtain
\begin{equation}
\delta\sigma /\sigma = {e\left[n(T,H)-n(T,0)\right]
\left[\mu_{p}(\nu)-\mu_{b}(\nu)\right]\over{\sigma(\nu,0)}}.
\end{equation}
All major features of the magnetooptical conductivity can be understood
with
this expression. First of all the magnetooptical effect is maximal
in the region of maximum variation of
the polaron density $n$ with the external magnetic field. It follows from our
equations (see below) that the effect is maximal close to $%
T_{c}$. The polaron mobility is larger than the bipolaron mobility at low
frequencies and smaller at high frequencies, Eq.(1). At the same time
one expects an increase of the polaron density with the magnetic field.
Hence the
magnetooptical effect should be positive in the low frequency region and
negative at high-frequencies.
To calculate the
effect we use the Hartree-Fock equations \cite{alebra,abopt}
accounting also for the external magnetic field as
\begin{eqnarray}
n&=& {\frac{t}{{2w}}}\ln \left[{\frac{1+2y \cosh({\frac{\sigma+h}{t}}) +
y^2%
}{{1+2ye^{-2w/t} \cosh({\frac{\sigma+h}{t}}) + y^2e^{-4w/t}}}}
\right],\label{eq:n} \\
m&=&{\frac{t}{{2w}}}\ln \left[\frac{{1+2y
e^{-w/t}\cosh({\frac{\sigma+h+w}{t}%
} ) + y^2 e^{-2w/t}}}{{1+2y e^{-w/t} \cosh({\frac{\sigma+h-w}{{t}}}) + y^2
e^{-2w/t}}}\right],  \label{eq:m} \\
\sigma&=&B_2 [(4h + m)/(2 t)] ,  \label{eq:s}
\end{eqnarray}
where
\begin{equation}
y= e^{-\delta/t} \left[{\frac{\sinh[(x-n)d/(2xt)]}{{%
\sinh[(x+n)d/(2xt)]}}}\right]^{1/2}.  \label{eq:y}
\end{equation}
Here $B_{S}(z) =\- [1+1/(2S)]\- \coth[(S+1/2)z]\- -[1/(2S)]\coth(z/2)$  is
the Brillouin function, $m$ and $\sigma$ are the relative magnetization of
polarons and of Mn,  respectively.  The reduced temperature is $%
t=2k_{B}T/(J_{pd}S)$,  the dimensionless binding energy $\delta =
\Delta/(J_{pd}S)$, and the dimensionless magnetic field is $2\mu_B
H/J_{pd}S$%
. These equations account for a finite polaron, $w = W/J_{pd}S $  and
bipolaron, $d = D/J_{pd}S $, widths of the energy level  distribution,
essential at low temperatures. We also assume here that  immobile
bipolarons
are localized by the impurities and there is no more than one bipolaron in
a single localized state (`single well-single particle' approximation \cite
{alebra3}).  Therefore, the total number of states in the bipolaron
(impurity) band is $x$.

The effect of magnetic field on optical conductivity is maximal near
the transition temperature, where the system exhibits a massive
spectral weight transfer between polaronic to bipolaronic peaks in
optical conductivity.
At low temperatures, $T<T_{c}$, when the system is almost
 polarized, $\sigma \approx 1$, the polaron density (\ref{eq:n})
depends on the field in the combination $1+{\frac{h}{t}}\tanh
({\frac{\sigma
}{t}})$. Since $h/t\ll 1$, the absolute change of the polaron density with
the field in this region is {\em always} small. The same is true of the
temperature range in the paramagnetic region far from the transition point.
However, close to $T_{c}$ the effect is very large as follows from
the main equations (\ref{eq:n})-(\ref{eq:y}).  Assuming a large carrier density
collapse, $n_{c}/x\ll 1$  we readily obtain from
Eq.~(\ref{eq:n}) (with $\sigma =0$) the density of polarons at the critical
temperature as
\begin{equation}
n_{c}(h)={\frac{t_{c}(h)}{w}}\exp [-\delta /t_{c}(h)]+O\left[ \left(
{h/t_{c}%
}\right) ^{2}\right] .  \label{eq:natc}
\end{equation}
Here we assume that $w\ll t_{c}$ (in the opposite limit, $w\gg t_{c}$, the
preexponential
factor will be 2 instead of $t_{c}/w)$. It formally looks as if we
have only a second-order effect of the field. However,
$t_{c}=t_{c}(h)$
is very sensitive to the field \cite{alebra2} and this leads to a strong
field dependence of the polaron density, and, consequently, to a colossal
magnetooptical effect.

\begin{figure}[t]
  \epsfxsize=3.4in
  \epsffile{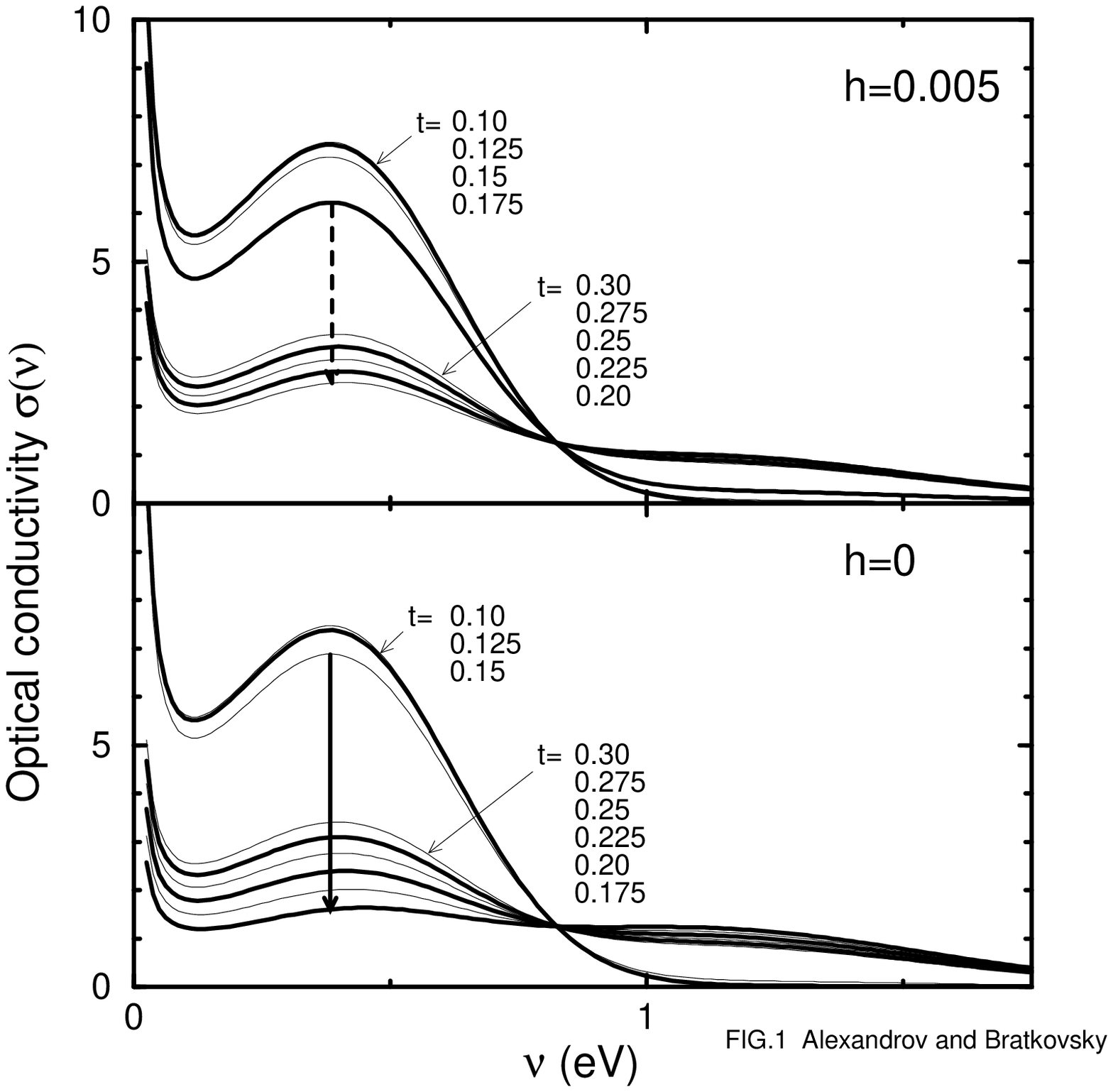 }
\caption{Optical conductivity $\sigma(\nu)$ for various temperatures
below and above the critical temperature $t_c=2k_BT_c/J_{pd}S$
($t_c=0.162$ for $\delta=\Delta/J_{pd}S=0.5$). Background contribution
to the absorption has been ignored [15]. The arrow indicates the
sudden drop in the 
polaronic contribution to the conductivity around $\nu_p=0.5$~eV when the
system is heated up through the critical temperature.
Top panel: magnetic field $h\equiv 2\mu_BH/J_{pd}S=0.005$; bottom
panel: $h=0$. 
    Note that the $\sigma(\nu)$ curve for $t=0.175$ raises from the
bottom of the lower manifold of curves at $h=0$ to join the upper
manifold of curves at $h=0.005$.
 For
La$_{0.75}$Ca$_{0.25}$MnO$_3$ we estimated $J_{pd}S=2250$~K [1], so
that the field $h=0.001$ corresponds to $H=1.67$~T.
}
\label{fig:sig}
\end{figure}

\begin{figure}[t]
  \epsfxsize=3.4in
 \epsffile{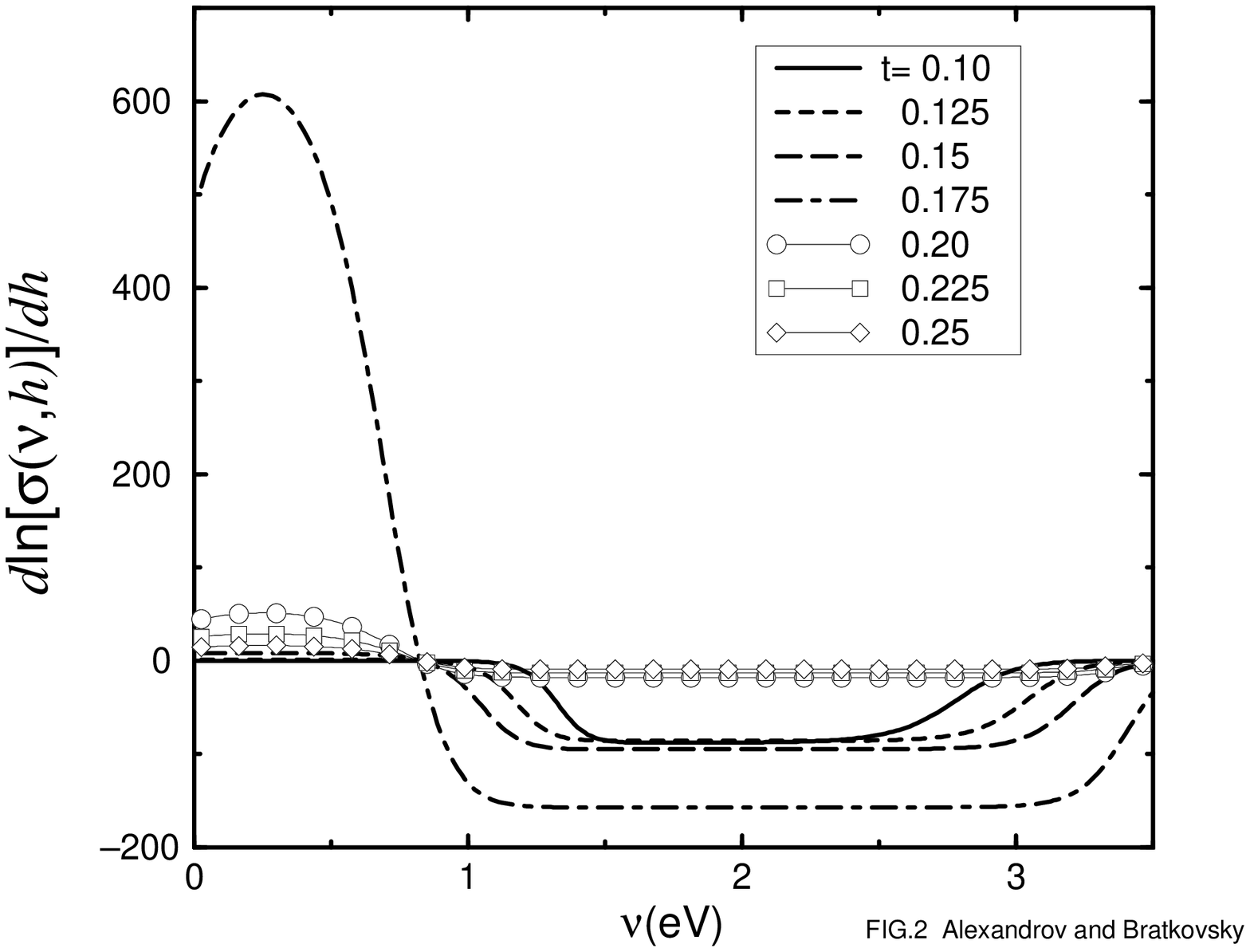 }
\caption{
Differential relative  magnetoconductivity  for the
system with the same parameters as in Fig.~\ref{fig:sig}.
Note the huge change in the conductivity close to the critical temperature
$t_c=2k_BT_c/J_{pd}S=0.162$. In a field $H=16$~T ($h\approx 0.01$), as
in experiment [16], the maximal effect is $(\delta\sigma/\sigma)_{max}=5.7$ at
$\nu=0.25$~eV and  $t=0.175$ (close to $T_c$).  
}
\label{fig:mo}
\end{figure}

To illustrate the point, we have calculated $\delta\sigma /\sigma$ for $\delta
=0.5$
($t_{c}=0.162$), with parameters $\nu _{p,b}$ and $\gamma _{p,b}$
given above, assuming that $d\ll 1$ (Figs.~\ref{fig:sig},
\ref{fig:mo}). With these
parameters the phase transition is discontinuous, and it is accompanied by
approximately an eight-fold decrease of the polaron density $n$ at the
transition. Consequently, the low frequency (polaronic)  peak at $\nu
_{p}=0.5$~eV
sharply reduces in height, as the spectral weight is suddenly
transferred to the bipolaronic peak at $\nu _{b}=1.24$~eV, for temperatures
varying between $t=0.15$ and $t=0.175$, through the critical temperature,
Fig.~\ref{fig:sig}. 
With further increase in temperature the density of
polarons  increases again since the bipolarons are subject to thermal
dissociation. As a result, we observe some re-entrant increase of the spectral
weight of the polaronic peak at lower frequencies at $T>T_{c}$.
It would be interesting to test this unusual temperature behavior of
conductivity experimentally.

In a wide frequency interval we obtain a several-fold change in the
magnetoconductivity
calculated for $h=0.005$  and temperature $t=0.175$, which
is somewhat higher than the critical temperature $t_c=0.162$.

The {\em positive} sign of the relative optical magnetoconductivity at
frequencies, 
where the dominant contribution comes from the polarons, immediately
follows
from our picture of current carrier density collapse. Indeed, the Zeeman
splitting of the polaron band pulls one of the spin subbands down in energy. A
fraction of the bipolarons, proportional to the magnetic field, will then be
destabilized and dissociate into polarons. This is accompanied by an
increase in the polaronic contribution to the 
conductivity and the positive sign of the magnetoconductance in the
low frequency region.
Below $T_c$ the relative changes in the conductivity are most pronounced in
the region $\nu \geq 1.5$~eV, where we have only tails of both polaron and
bipolaron contributions. These changes are exaggerated  because $%
\sigma(\nu,0)$ itself is small in this region. Above $T_c$ the
magnetooptical effect is mostly pronounced at low frequencies $\nu
\leq 1$~eV (Fig.~\ref{fig:mo}). Our theoretical prediction of a colossal and positive
relative magnetoconductivity near $T_c$ 
is in agreement with recent measurements \cite{boris}.

The spectral weight transfer is  usually measured by the
effective optical number of carriers
\begin{equation}
N_{eff}(\nu _{c})={\frac{2m_{0}}{\pi e^{2}N}}\int_{0}^{\nu _{c}}d\nu 
\sigma (\nu ),  \label{eq:neff}
\end{equation}
where $m_{0}$ is the bare electron mass, $N$ is the number of, for
instance, dopant 
atoms per unit volume, $\nu _{c}$ is the cutoff energy, which spans
a low-frequency 
region (e.g. $\nu_c=0.5$~eV\cite{kim}, 0.8~eV\cite{ish}). According to
the sum rule for 
conductivity, $N_{eff}(\infty )$  equals the number of
carriers
in the system. By using a finite-frequency cutoff, one can estimate
the integral weight
transfer from a given interval of energies. Typical behavior of doped
manganites is that  $N_{eff}$ for mid-IR gradually diminishes with
elevated temperature in the ferromagnetic phase until the systems
approaches a critical temperature, where the weight drops.
With the mobilities, Eq.~(1), calculated within a (nonconserving)
saddle-point approximation one is unable to quantify the integral spectral
weight transfer. However, its experimental
 behavior is in qualitative agreement  with our notion of the current
carrier density collapse as described above.

In conclusion, we have developed the theory of the magnetooptical
conductivity in doped magnetic charge-transfer insulators with a strong
electron-phonon interaction. We have found a colossal magnetooptical effect
close to the critical temperature, positive at low frequencies and
negative at high frequencies. The conductivity increase in the field is
several-fold at low frequencies near the phase transition in
accordance with that observed 
experimentally\cite{boris}. Our theory of doped manganites is largely
independent of the symmetry of the relevant electron orbitals and of the type
of electron-phonon interaction. Its key feature is the pairing of
carriers in the paramagnetic phase due to the lattice distortion and
their unpairing in the ferromagnetic phase due to the exchange pair-breaking
interaction, the latter being
reminiscent of the magnetic pair-breaking in superconductors with
magnetic impurities. It would be interesting to check the  predicted
strong temperature dependence of the magnetoconductivity
experimentally on various CMR systems.

We would like to acknowledge useful discussions with D.N. Basov
pertaining to peculiarities of optical properties of strongly correlated
systems. We are grateful to
G. Aeppli, A.R. Bishop, D.S. Dessau, M.F. Hundley, K.M. Krishnan, A.P.
Ramirez, R.S. Williams, and Guo-meng Zhao for helpful discussions. ASA
acknowledges support from the Quantum Structures Research Initiative and the
External Research Program of Hewlett-Packard Laboratories (Palo Alto).




\begin{references}
\bibitem[*]{email}  Email asa21@cus.cam.ac.uk

\bibitem[\dagger]{email}  Email alexb@hpl.hp.com


\bibitem{alebra}   A.S. Alexandrov and A.M. Bratkovsky, Phys. Rev. Lett. $%
{\bf 82}$, 141  (1999).

\bibitem{alebra2}   A.S. Alexandrov and A.M. Bratkovsky, J. Phys. Condens.
Matter {\bf \ 11}, 1989 (1999).

\bibitem{van}   G.H. Jonker and J.H. Van Santen, Physica (Utrecht) ${\bf
16}$%
, 337 (1950);  J.H. Van Santen and G.H. Jonker, Physica (Utrecht) ${\bf
16}$%
, 599 (1950).

\bibitem{helmolt}   R. von Helmolt $et$ $al.$, Phys. Rev. Lett. ${\bf 71}$,
2331 (1993).

\bibitem{jin}   S. Jin $et$ $al.$, Science ${\bf 264}$, 413 (1994).


\bibitem{mul}   Guo-meng Zhao {\em et al.}, Nature {\bf 381}, 676 (1996).

\bibitem{franck}  J.P. Franck {\em et al.}, \prb {\bf 58}, 5189 (1998).

\bibitem{tun}   A. Biswas $et$ $al.$, cond-mat/9806084.

\bibitem{ram}   A.P. Ramirez, J. Phys. Condens. Matter {\bf 9}, 8171
(1997).

\bibitem{sch}   P. Schiffer $et$ $al.$, Phys. Rev. Lett. {\bf 75}, 3336
(1995).

\bibitem{oki1}   Y.Okimoto {\em et al.}, Phys. Rev. Lett. {\bf 75}, 109
(1995).

\bibitem{oki2}   Y.Okimoto {\em et al.}, Phys. Rev. B{\bf 55}, 4206 (1997).

\bibitem{kim}   K.H. Kim, J.H. Yung, and T.W. Noh, Phys. Rev. Lett. {\bf
81}%
, 1517 (1998).

\bibitem{ish}   T. Ishikawa $et$ $al.$, Phys. Rev. B {\bf 57}, R8079
(1998).

\bibitem{abopt}  A.S.Alexandrov~and~A.M.Brat\-kov\-sky, co\-nd\--mat/\-9901340\-
(1999).

\bibitem{boris}  A.V. Boris {\em et al.}, \prb {\bf 59}, 697 (1999).

\bibitem{des}   D.S. Dessau {\em et al.}, Phys. Rev. Lett. {\bf 81}, 192
(1998).

\bibitem{gud}   J.-S. Zhou $et$ $al.$, Phys. Rev. Lett. ${\bf 79}$, 3234
(1997).

\bibitem{bri}   H. B\"ottger and V.V. Bryksin, {\it Hopping Conduction in
Solids}  (Academic-Verlag, Berlin, 1985).

\bibitem{mah}   G.D. Mahan, {\it Many-Particle Physics} (Plenum, New York,
1990).

\bibitem{alemot}   A.S. Alexandrov and N.F. Mott, {\em Polarons and
Bipolarons}  (World Scientific, Singapore, 1995).

\bibitem{alebra3}   A.S. Alexandrov, A.M. Bratkovsky, and N.F. Mott, Phys.
Rev. Lett. {\bf 72},  1734 (1994).
\end{references}
\end{document}